\documentclass[conference]{IEEEtran}
\IEEEoverridecommandlockouts
\usepackage[a4paper,top=0.75in,bottom=0.8in,left=0.75in,right=0.8in]{geometry}
\usepackage{url}

\usepackage{cite}
\usepackage{amsmath,amssymb,amsfonts}
\usepackage{algorithm}
\usepackage{algpseudocode}
\usepackage{graphicx}
\usepackage{textcomp}
\usepackage{xcolor}
\usepackage{booktabs}
\usepackage{graphicx}
\usepackage{comment}
\usepackage{colortbl}
\usepackage{multirow}

\renewcommand{\arraystretch}{1.3}

\def\BibTeX{{\rm B\kern-.05em{\sc i\kern-.025em b}\kern-.08em
    T\kern-.1667em\lower.7ex\hbox{E}\kern-.125emX}}
\begin{document}

\title{Is Mamba Reliable for Medical Imaging?\\
{\footnotesize }
\thanks{}
}

\author{\IEEEauthorblockN{
Banafsheh Saber Latibari\textsuperscript{1},
Najmeh Nazari\textsuperscript{2},
Daniel Brignac\textsuperscript{1},
Hossein Sayadi\textsuperscript{3},
\\ Houman Homayoun\textsuperscript{2},
Abhijit Mahalanobis\textsuperscript{1}
}
\IEEEauthorblockA{
\textsuperscript{1}Department of Electrical and Computer Engineering, University of Arizona, Tucson, AZ, USA \\
\textsuperscript{2}Department of Electrical and Computer Engineering, University of California, Davis, CA, USA\\
\textsuperscript{3}Department of  Computer Engineering and Computer Science, California State University, Long Beach, \\ Long Beach, CA, USA \\
Emails: \{banafsheh, dbrignac, amahalan\}@arizona.edu , \{nnazari, hhomayoun\}@ucdavis.edu},  hossein.sayadi@csulb.edu
}

\maketitle

\begin{abstract}
State-space models like Mamba offer linear-time sequence processing and low memory, making them attractive for medical imaging. However, their robustness under realistic software and hardware threat models remains underexplored. This paper evaluates Mamba on multiple MedMNIST classification benchmarks under input-level attacks, including white-box adversarial perturbations (FGSM/PGD), occlusion-based PatchDrop, and common acquisition corruptions (Gaussian noise and defocus blur) as well as hardware-inspired fault attacks emulated in software via targeted and random bit-flip injections into weights and activations. We profile vulnerabilities and quantify impacts on accuracy indicating that defenses are needed for deployment.
\end{abstract}


\section{Introduction}

\begin{figure}
    \centering
    \includegraphics[width=\linewidth]{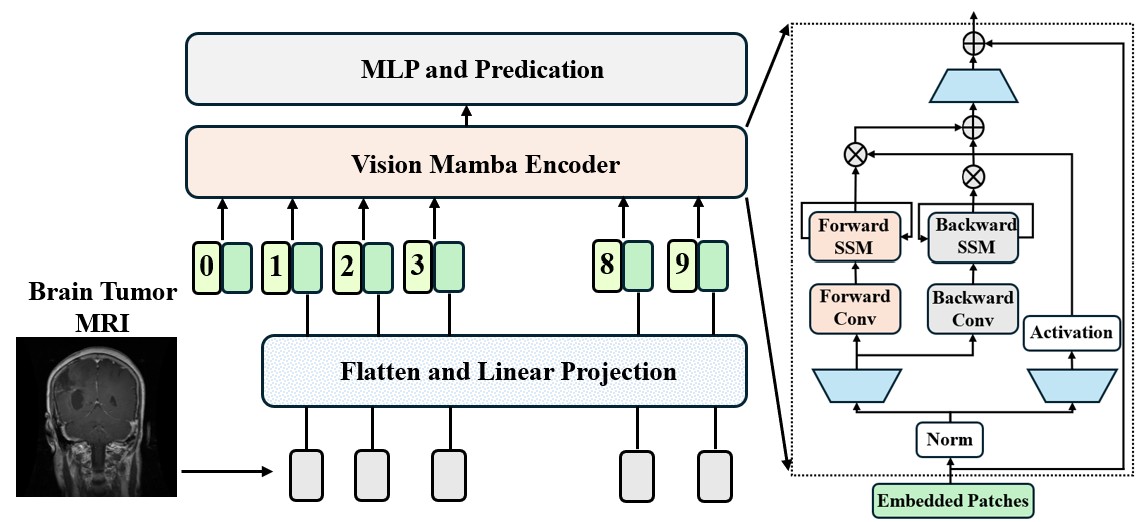}
    \caption{Architecture of Vision Mamba}
    \label{Mamba}
\end{figure}

Transformers are widely used in various applications; however, the quadratic computational cost of the attention mechanism limits their widespread adoption \cite{saber2024iret, latibari2025optimizing}. Recently, a new model from the state-space model family, called Mamba \cite{gu2024mambalineartimesequencemodeling}, has been introduced, which has linear computational complexity. Mamba architecture is adapted for computer vision applications, and Vision Mamba (ViM) is introduced by Zhu et al. \cite{zhu2024vision}, which uses bidirectional SSM by combining convolution with S6 in both forward and backward directions. Fig. \ref{Mamba} shows the overview of ViM.

AI is transforming healthcare, and one of the main domains that benefits from it is medical imaging. Several architectures have been tested for different medical imaging tasks, including quality enhancement, segmentation, detection, and classification. Mamba has recently attracted significant attention in this domain \cite{bansal2024comprehensive}. However, the reliability and security of these models need attention. In other words, we need to answer the question, is Mamba reliable for medical imaging? Previously, it has been demonstrated that Transformers pose security concerns at both the software and hardware levels for several applications, including medical imaging \cite{latibari2024transformers, nazari2025faraccel, latibari2025hammering}.

In this work, we will evaluate the reliability of Mamba and perform hardware and software threats to its attack surface. This work has the following contributions:

\begin{itemize}
 \item We provide a comprehensive robustness evaluation of Mamba on multiple MedMNIST datasets under \textbf{input-level attacks}, including white-box adversarial perturbations (FGSM/PGD) \cite{mirzaeian2022adaptive}, information drop (PatchDrop), and realistic corruption shifts (Gaussian noise and defocus blur).
    \item We introduce \textbf{Med-Mamba-Hammer}, a hardware-inspired fault analysis framework that evaluates \textbf{random}, \textbf{layer-wise}, and \textbf{worst-case} bit-flip injection, capturing potential memory fault behaviors such as DRAM-induced errors.
    \item We quantify the relative sensitivity of different model components to faults and show that early feature extraction and SSM-related modules are more vulnerable than later stages, highlighting key reliability bottlenecks.
    \item We demonstrate that small fault budgets can lead to severe accuracy degradation, including extreme worst-case failures where a single high-impact bit flip can collapse performance to near-random accuracy.
\end{itemize}

\begin{figure*}
    \centering
    \includegraphics[width=0.7\textwidth]{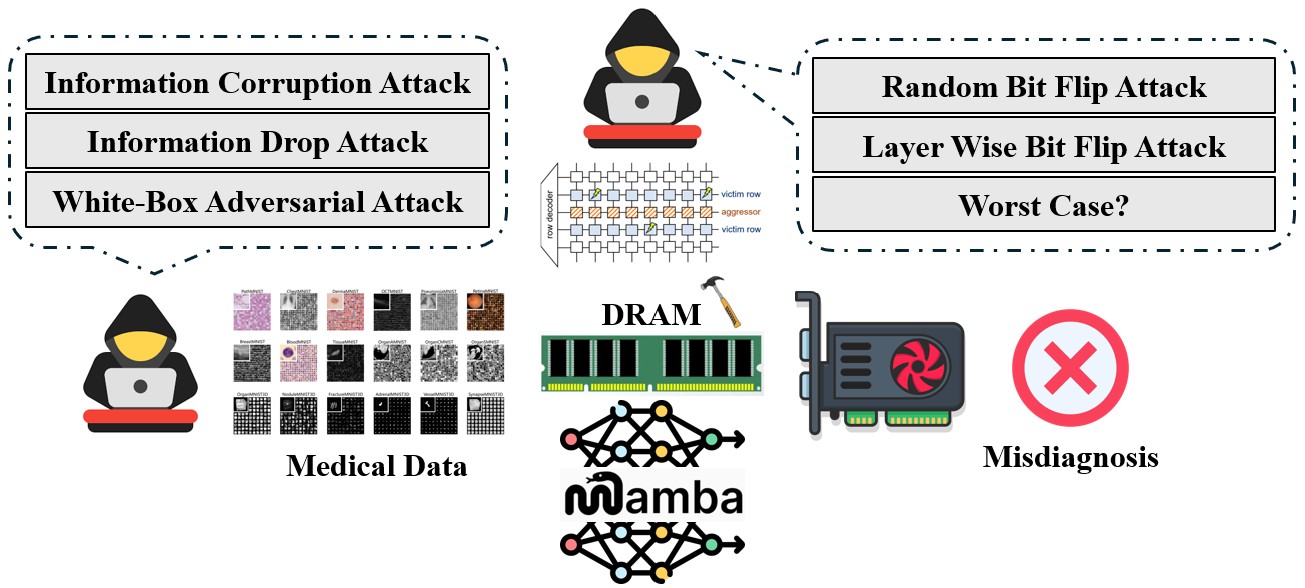}
    \caption{Overview of the threat model and evaluation pipeline for medical image classification. We consider input-level attacks including information corruption, information drop, and white-box adversarial perturbations, as well as hardware-level fault attacks such as random and layer-wise bit flips (e.g., Rowhammer-induced DRAM faults). The diagram highlights how these disturbances can propagate through the model and potentially lead to incorrect clinical predictions (misdiagnosis).}
    \label{fig:placeholder}
\end{figure*}

\section{Background}

\textbf{Mamba in Medical Imaging-}
Mamba has performed impressively in the field of medical imaging. In medical image classification, nnMamba \cite{gong2025nnmamba} combines the strengths of CNNs and SSMs, introducing channel scaling and channel-sequential learning. Medmamba \cite{yue2024medmamba} uses a patch embedding layer, SS-Conv-SSM blocks, and patch merging layers, with dual branches that integrate convolutional and SSM features through a 2D selective scan for richer feature extraction. SegMamba \cite{xing2024segmamba} is a U-Net–like architecture trained on a new dataset of 500 expert-annotated 3D CT scans. H-Vmunet \cite{wu2025h} is a transformer-inspired U-Net that uses the H-VSS module, replacing VSS with an order-dependent H-SS2D channel-splitting mechanism. It stacks H-VSS modules across encoder and decoder stages, uses shared CAB and SAB for feature fusion, and upsamples with bilinear interpolation. FDVM-Net \cite{zheng2024fd} and MambaMIR \cite{huang2024mambamirarbitrarymaskedmambajoint} are Mamba-based models for image restoration. FDVM-Net improves endoscopic exposure correction using frequency-domain reconstruction with a dual-path Mamba–CNN design and frequency-domain cross-attention. MambaMIR targets medical image reconstruction and uncertainty estimation through an AMSS block with Monte Carlo dropout and Mamba-inspired components stabilized by residual connections. MambaMorph \cite{guo2025mambamorphmambabasedframeworkmedical} is a multi-modality deformable registration framework that uses a Mamba-based module and a lightweight feature extractor for efficient long-range correspondence modeling.

\textbf{Security in Mamba-based Medical Imaging-} Ensuring the security and reliability of an AI model before using it in a critical application has significant importance. However, there is limited work on evaluating the security of Mamba for medical imaging, and there is no previous work from a hardware security perspective on Mamba. Authors proposed Mamba-Fusion \cite{jabbar2025mamba}, a privacy-preserving multi-modal disease prediction framework that uses hierarchical federated learning and an MoE–LSTM architecture to improve scalability and temporal integration. By combining differential privacy and secure aggregation, it achieves 92.4\% accuracy with low privacy leakage and communication cost, outperforming conventional FL methods. In \cite{malik2024evaluating}, the authors investigate the adversarial robustness of volumetric medical segmentation models, including Convolutional, Transformer, and Mamba-based architectures, across four datasets under both white-box and black-box attacks. They find that frequency-based attacks transfer better in black-box settings, Transformer models are more robust than convolution-based ones, and Mamba-based models are the most vulnerable. They also show that large-scale training improves robustness against adversarial attacks.

\section{Med-Mamba-Adv: Input-level perturbation for Mamba-based Medical Imaging}
This section evaluates how robust a Mamba-based medical imaging model is when the input is intentionally or naturally degraded. We focus on three complementary robustness axes: (1) white-box adversarial perturbations (FGSM/PGD), (2) information drop/occlusion (missing or masked content), and (3) common corruptions (noise/blurring).

\subsection{White box Adversarial Attack} We evaluated the robustness of models by performing a range of gradient-based attacks, such as FGSM and PGD. We assume a white-box attacker who knows the model parameters and can compute gradients. 
Let $f_\theta(\cdot)$ denote the trained model and $\mathcal{L}(\cdot,\cdot)$ the task loss (e.g., cross-entropy for classification). We primarily consider a white-box attacker that has access to $\theta$ and crafts a perturbed input $x^{adv}$ within an $\ell_\infty$-bounded ball around the clean input $x$:
\begin{equation}
x^{adv} \in \mathcal{B}_\infty(x,\epsilon) := \{x' : \|x'-x\|_\infty \le \epsilon\}, \quad x^{adv}\in[0,1].
\end{equation}

\paragraph{FGSM}
Fast Gradient Sign Method perturbs the input once in the direction of the gradient sign:
\begin{equation}
x^{adv} = \mathrm{clip}_{[0,1]}\big(x + \epsilon \cdot \mathrm{sign}(\nabla_x \mathcal{L}(f_\theta(x),y))\big).
\end{equation}

\paragraph{PGD}
Projected Gradient Descent is a multi-step, first-order attack that iterates BIM with a random start (optional) and projection:
\begin{equation}
x_{t+1}^{adv}=\Pi_{\mathcal{B}_\infty(x,\epsilon)}\Big(x_t^{adv}+\alpha \cdot \mathrm{sign}(\nabla_{x}\mathcal{L}(f_\theta(x_t^{adv}),y))\Big).
\end{equation}
As a default setting, we follow a widely used configuration for $\ell_\infty$-PGD, using $\epsilon=1/255$ with $T=20$ steps (and sweep $\epsilon$ where appropriate).

\subsection{Information Drop Robustness}
Clinical images frequently contain missing, corrupted, or occluded regions due to partial field-of-view, sensor dropout, motion-induced artifacts, masking, or tool/implant occlusions. To model this setting, we define an information drop robustness protocol that systematically removes localized evidence from the input while preserving the remaining content.

Given an input image $x$, we divide it into $N$ non-overlapping patches $\{p_i\}_{i=1}^{N}$. We randomly choose a fraction $r$ of patches to remove, i.e., a set $\mathcal{S}$ with $|\mathcal{S}|=\lfloor rN\rfloor$. We form the dropped image $\tilde{x}$ by replacing the selected patches with a baseline value $b$ (e.g., $0$ after normalization):
\begin{equation}
\tilde{x}[p_i] =
\begin{cases}
b, & i \in \mathcal{S},\\
x[p_i], & \text{otherwise}.
\end{cases}
\end{equation}
This patch-drop operation measures how well the model tolerates missing local information, as commonly occurs with occlusions and data loss in clinical images.

\subsection{Information Corruption Robustness}
While adversarial perturbations capture worst-case, optimization-driven shifts, medical images are often degraded by \emph{natural} acquisition and processing artifacts (e.g., sensor noise, patient motion, defocus, reconstruction smoothing, and compression). To model these clinically realistic shifts, we define an information corruption robustness setting in which the input image is transformed by a corruption operator before being processed by the model.

Formally, given a clean input image $x$, we construct a corrupted input
\begin{equation}
\tilde{x} = \mathcal{C}(x; s),
\end{equation}
where $\mathcal{C}(\cdot)$ denotes a corruption function and $s \in \{1,\dots,S\}$ controls the \textbf{severity level} (from mild to severe). Unlike adversarial attacks, $\mathcal{C}$ is not gradient-based; instead, it represents plausible degradations that preserve the underlying semantic content while reducing the diagnostic signal available to the model.

In this work, we focus on two common corruption families:

\textbf{Gaussian noise.} We inject additive Gaussian noise to mimic stochastic acquisition noise and low-signal conditions:
\begin{equation}
\tilde{x} = \mathrm{clip}_{[0,1]}\big(x + \eta\big), \quad \eta \sim \mathcal{N}(0,\sigma_s^2),
\end{equation}
where $\sigma_s$ increases with severity $s$ and clipping enforces valid intensity ranges.

\textbf{Blurring.} We apply blur to simulate loss of high-frequency anatomical detail caused by motion, defocus, or smoothing:
\begin{equation}
\tilde{x} = x * k_s,
\end{equation}
where $k_s$ is a blur kernel (e.g., Gaussian/defocus) whose spread increases with $s$.

This corruption-based threat model complements adversarial robustness by evaluating stability under realistic, deployment-relevant image quality shifts.

\begin{algorithm}[t]
\caption{Random (Global) Bit-Flip Evaluation}
\label{alg:random_bitflip}
\begin{algorithmic}[1]

\Statex \textbf{Input:} Trained weights $W$, test loader $\mathcal{D}_{test}$, bit budgets $\mathcal{K}$,
number of trials $T$, bit region $R$
\Statex \textbf{Output:} Mean and std accuracy for each $K \in \mathcal{K}$

\State $A_{\text{base}} \gets \mathrm{Acc}(\mathrm{Model}(W), \mathcal{D}_{test})$

\For{$K \in \mathcal{K}$}
    \State $\mathcal{A} \gets [\,]$
    \For{$t = 1 \ \textbf{to} \ T$}
        \State $W' \gets \mathrm{InjectBitFlips}(W,\;K,\;R,\;\text{seed}=1234+t)$
        \State $M \gets \mathrm{Model}(W')$
        \State $a \gets \mathrm{Acc}(M,\mathcal{D}_{test})$
        \State $\mathcal{A}.\mathrm{append}(a)$
    \EndFor
    \State $\mu_K \gets \mathrm{mean}(\mathcal{A})$
    \State $\sigma_K \gets \mathrm{std}(\mathcal{A})$
    \State \textbf{report} $(K,\mu_K,\sigma_K)$
\EndFor

\end{algorithmic}
\end{algorithm}

\begin{algorithm}[t]
\caption{Layer-wise Bit-Flip Evaluation}
\label{alg:layerwise_bitflip}
\begin{algorithmic}[1]

\Statex \textbf{Input:} trained weights $W$, test loader $\mathcal{D}_{test}$, bit budgets $\mathcal{K}$,
trials $T$, bit region $R$, layer groups $\mathcal{G}$ (each provides key-filter $\mathrm{filter}_g$)
\Statex \textbf{Output:} mean and std accuracy for each group $g$ and each $K \in \mathcal{K}$

\State $A_{\text{base}} \gets \mathrm{Acc}(\mathrm{Model}(W), \mathcal{D}_{test})$

\For{$K \in \mathcal{K}$}
    \For{$(g,\mathrm{filter}_g) \in \mathcal{G}$}
        \State $\mathcal{A} \gets [\,]$
        \For{$t = 1$ \textbf{to} $T$}
            \State $W' \gets \mathrm{InjectBitFlips}(W,\;K,\;R,\;\text{seed}=1234+t,$
            \Statex \hspace{\algorithmicindent} $\text{key\_filter}=\mathrm{filter}_g)$
            \State $M \gets \mathrm{Model}(W')$
            \State $a \gets \mathrm{Acc}(M,\mathcal{D}_{test})$
            \State $\mathcal{A}.\mathrm{append}(a)$
        \EndFor
        \State $\mu_{K,g} \gets \mathrm{mean}(\mathcal{A})$
        \State $\sigma_{K,g} \gets \mathrm{std}(\mathcal{A})$
        \State \textbf{report} $(K,g,\mu_{K,g},\sigma_{K,g})$
    \EndFor
\EndFor

\end{algorithmic}
\end{algorithm}

\begin{algorithm}[t]
\caption{Worst-Case Bit-Flip Search (Random-Search Adversary)}
\label{alg:worstcase_bitflip}
\begin{algorithmic}[1]

\Statex \textbf{Input:} trained weights $W$, test loader $\mathcal{D}_{test}$, bit budgets $\mathcal{K}$,
bit region $R$, search iterations $N$, fast-eval batches $B$, optional key-filter $\mathrm{filter}$
\Statex \textbf{Output:} worst-case seed and full-test accuracy for each $K \in \mathcal{K}$

\State $A_{\text{base}} \gets \mathrm{Acc}(\mathrm{Model}(W), \mathcal{D}_{test})$

\For{$K \in \mathcal{K}$}
    \State $a_{\min} \gets +\infty$
    \State $s^\star \gets \varnothing$
    \State $W^\star \gets \varnothing$

    \For{$i = 1$ \textbf{to} $N$}
        \State $s \gets 9000 + i$
        \State $W' \gets \mathrm{InjectBitFlips}(W,\;K,\;R,\;\text{seed}=s,$
        \Statex \hspace{\algorithmicindent} $\text{key\_filter}=\mathrm{filter})$
        \State $M \gets \mathrm{Model}(W')$
        \State $a_{\text{fast}} \gets \mathrm{Acc}(M,\mathcal{D}_{test}[1{:}B])$
        \Statex \hspace{\algorithmicindent} \Comment{evaluate on first $B$ batches}

        \If{$a_{\text{fast}} < a_{\min}$}
            \State $a_{\min} \gets a_{\text{fast}}$
            \State $s^\star \gets s$
            \State $W^\star \gets W'$
        \EndIf
    \EndFor

    \State $M^\star \gets \mathrm{Model}(W^\star)$
    \State $a_{\text{full}} \gets \mathrm{Acc}(M^\star,\mathcal{D}_{test})$
    \Statex \hspace{\algorithmicindent} \Comment{full evaluation of worst candidate}
    \State \textbf{report} $(K,\;s^\star,\;a_{\min},\;a_{\text{full}})$
\EndFor

\end{algorithmic}
\end{algorithm}

\begin{table}[]
\centering
\caption{Model accuracy (\%) on MedMNIST datasets under clean, FGSM, and PGD ($\epsilon = 1/255, 20$ steps).}
\label{Wattack}
\resizebox{0.7\columnwidth}{!}{%
\begin{tabular}{@{}|c|c|c|c|@{}}
\toprule
\rowcolor[HTML]{EFEFEF} 
Dataset & Clean Acc & FGSM Acc & PGD Acc \\ \midrule
PathMNIST & 89.73 & 26.31 & 10.56 \\ \midrule
DermaMNIST & 79.35 & 20.35 & 3.89 \\ \midrule
OCTMNIST & 91.8 & 41.2 & 20.6  \\ \midrule
\multicolumn{1}{|l|}{PneumoniaMNIST} & 93.27 & 13.46 & 0 \\ \midrule
RetinaMNIST & 51.0 & 22.5 & 4.25 \\ \midrule
BreastMNIST & 88.46 & 20.51 & 8.33 \\ \midrule
BloodMNIST & 97.63 & 58.02 & 48.29 \\ \midrule
OrganAMNIST & 89.19 & 57.85 & 52.74 \\ \midrule
OrganCMNIST & 89.02 & 51.90 & 45.20 \\ \midrule
OrganSMNIST & 73.39 & 32.34 & 24.84 \\ \bottomrule
\end{tabular}%
}
\end{table}

\begin{table}[]
\centering
\caption{Robustness evaluation using PatchDrop showing classification accuracy (\%) across MedMNIST datasets under increasing patch removal ratios.}
\label{drop}
\resizebox{\columnwidth}{!}{%
\begin{tabular}{@{}|c|c|c|c|c|c|c|c|@{}}
\toprule
\rowcolor[HTML]{EFEFEF} 
Dataset & 0\% & 6.25\% & 18.75\% & 25\% & 37.5\% & 50\% & 56.25\%  \\ \midrule
PathMNIST & 89.72 & 91.8 & 80.86 & 72.81 & 56.73 & 45.64 & 38.95 \\ \midrule
DermaMNIST & 79.35 & 72.77 & 55.81 & 39.75 & 20.70 & 17.85 & 17.75  \\ \midrule
OCTMNIST & 91.8 & 85.1 & 71.1 & 67.9 & 61.2 & 47.1 & 42.4  \\ \midrule
PneumoniaMNIST & 93.27 & 91.03 & 85.26 & 78.53 & 62.82 & 56.73 & 43.11  \\ \midrule
RetinaMNIST & 51.0 & 51.5 & 51.75 & 51.5 & 50.5 & 45.75 & 44.75 \\ \midrule
BreastMNIST & 88.46 & 86.54 & 81.41 & 78.85 & 76.28 & 69.87 & 73.07  \\ \midrule
BloodMNIST & 97.63 & 79.19 & 32.88 & 22.54 & 11.34 & 9.24 & 9.12 \\ \midrule
OrganAMNIST & 89.19 & 86.67 & 77.62 & 71.04 & 57.56 & 42.92 & 36.15  \\ \midrule
OrganCMNIST & 89.02 & 85.82 & 76.92 & 70.59 & 58.01 & 48.26 & 43.62 \\ \midrule
OrganSMNIST & 73.39 & 66.59 & 53.92 & 48.09 & 32.42 & 19.97 & 16.03  \\ \bottomrule
\end{tabular}%
}
\end{table}

\section{Med-Mamba-Hammer: Hardware-Level bit-flip attack for Mamba-based Medical Imaging}

Deep medical imaging models deployed on edge devices and hospital infrastructure can be exposed to \emph{hardware-induced} faults, such as transient memory errors, aging effects, or disturbance errors that manifest as unintended \textbf{bit flips} in stored parameters or intermediate values. Unlike input-space perturbations, these faults act \emph{inside} the inference pipeline and may cause silent but significant performance degradation. We term this robustness setting \textbf{Med-Mamba-Hammer}, a hardware-level ``bit-flip'' threat model tailored to Mamba-based medical imaging systems.

\subsection{Threat Model}
We consider an attacker that can induce a small number of bit flips during inference or model storage, affecting:
(i) \textbf{model weights} (e.g., parameters stored in DRAM/Flash), and/or
(ii) \textbf{activations} (intermediate feature maps stored in memory during inference).
We assume the attacker does not modify the input image directly. The goal is to reduce diagnostic performance (untargeted) or induce specific incorrect outputs (targeted), while flipping only a limited number of bits.

\subsection{Bit-Flip Injection Model}
Let $\theta$ denote the model parameters and let $\tilde{\theta}$ be the corrupted parameters after bit flips. We model a bit flip as an operator $\mathcal{B}(\cdot)$ applied to the binary representation of a value:
\begin{equation}
\tilde{\theta} = \mathcal{B}(\theta; \mathcal{I}),
\end{equation}
where $\mathcal{I}$ is the set of flipped bit indices across the parameter tensor(s). The number of flips is constrained by a budget $K = |\mathcal{I}|$. For a parameter element $\theta_j$ represented with $m$ bits, flipping bit $k$ produces:
\begin{equation}
\tilde{\theta}_j = \theta_j \oplus (1 \ll k),
\end{equation}
where $\oplus$ denotes bitwise XOR. In floating-point, flips in the sign/exponent bits can induce large numerical changes, while mantissa flips typically cause smaller perturbations.

\subsection{Attack Variants}
To characterize robustness under different fault patterns, we evaluate multiple bit-flip strategies:

\textbf{Random Bit-Flip.} Bits are flipped uniformly at random under a fixed budget $K$. This approximates non-adversarial hardware faults.

\textbf{Layer-wise Bit-Flip.} Bit flips are restricted to a specific layer or module (e.g., early feature extraction, SSM blocks, or classifier/decoder heads) to measure \emph{sensitivity} across the network.

\textbf{Worst-Case Bit-Flip.} We report the most damaging configuration within a constrained search space (e.g., flipping a small number of high-impact bits), representing a stronger adversary than purely random faults. This provides an upper bound on vulnerability under budgeted bit corruption.

\begin{table*}[]
\centering
\caption{Accuracy (\%) under Gaussian noise and defocus blur at different intensity levels.}
\label{noise}
\resizebox{0.7\textwidth}{!}{%
\begin{tabular}{|c|c|cc|cc|cc|cc|cc|}
\toprule
\rowcolor[HTML]{EFEFEF} 
\cellcolor[HTML]{EFEFEF} & \cellcolor[HTML]{EFEFEF} & \multicolumn{2}{c|}{\cellcolor[HTML]{EFEFEF}Intensity 1} & \multicolumn{2}{c|}{\cellcolor[HTML]{EFEFEF}Intensity 2} & \multicolumn{2}{c|}{\cellcolor[HTML]{EFEFEF}Intensity 3} & \multicolumn{2}{c|}{\cellcolor[HTML]{EFEFEF}Intensity 4} & \multicolumn{2}{c|}{\cellcolor[HTML]{EFEFEF}Intensity 5} \\ \cline{3-12} 
\rowcolor[HTML]{EFEFEF} 
\multirow{-2}{*}{\cellcolor[HTML]{EFEFEF}Dataset} & \multirow{-2}{*}{\cellcolor[HTML]{EFEFEF}Clean Acc} & \multicolumn{1}{c|}{\cellcolor[HTML]{EFEFEF}N} & B & \multicolumn{1}{c|}{\cellcolor[HTML]{EFEFEF}N} & B & \multicolumn{1}{c|}{\cellcolor[HTML]{EFEFEF}N} & B & \multicolumn{1}{c|}{\cellcolor[HTML]{EFEFEF}N} & B & \multicolumn{1}{c|}{\cellcolor[HTML]{EFEFEF}N} & B \\ \midrule
PathMNIST & 89.72 & \multicolumn{1}{c|}{61.99} & 81.89 & \multicolumn{1}{c|}{33.20} & 72.94 & \multicolumn{1}{c|}{23.22} & 57.73 & \multicolumn{1}{c|}{9.04} & 45.89 & \multicolumn{1}{c|}{6.13} & 40.89 \\ \midrule
DermaMNIST & 79.35 & \multicolumn{1}{c|}{70.47} & 78.80 & \multicolumn{1}{c|}{66.73} & 78.06 & \multicolumn{1}{c|}{66.93} & 76.51 & \multicolumn{1}{c|}{66.88} & 75.86 & \multicolumn{1}{c|}{66.88} & 74.96 \\ \midrule
OCTMNIST & 91.80 & \multicolumn{1}{c|}{84.40} & 83.90 & \multicolumn{1}{c|}{71.10} & 71.50 & \multicolumn{1}{c|}{54.70} & 57.00 & \multicolumn{1}{c|}{26.90} & 48.70 & \multicolumn{1}{c|}{25.00} & 42.60 \\ \midrule
PneumoniaMNIST & 93.27 & \multicolumn{1}{c|}{62.98} & 61.54 & \multicolumn{1}{c|}{37.50} & 58.65 & \multicolumn{1}{c|}{37.50} & 63.30 & \multicolumn{1}{c|}{37.50} & 73.40 & \multicolumn{1}{c|}{37.50} & 82.37 \\ \midrule
RetinaMNIST & 51.00 & \multicolumn{1}{c|}{45.75} & 50.75 & \multicolumn{1}{c|}{44.50} & 51.50 & \multicolumn{1}{c|}{43.50} & 51.00 & \multicolumn{1}{c|}{43.50} & 50.75 & \multicolumn{1}{c|}{43.50} & 51.00 \\ \midrule
BreastMNIST & 88.46 & \multicolumn{1}{c|}{74.36} & 86.54 & \multicolumn{1}{c|}{74.36} & 85.90 & \multicolumn{1}{c|}{73.10} & 78.85 & \multicolumn{1}{c|}{73.10} & 73.10 & \multicolumn{1}{c|}{73.10} & 62.18 \\ \midrule
BloodMNIST & 97.63 & \multicolumn{1}{c|}{65.44} & 85.53 & \multicolumn{1}{c|}{42.97} & 73.14 & \multicolumn{1}{c|}{25.23} & 59.16 & \multicolumn{1}{c|}{18.27} & 53.10 & \multicolumn{1}{c|}{5.50} & 44.00 \\ \midrule
OrganAMNIST & 89.19 & \multicolumn{1}{c|}{85.04} & 90.29 & \multicolumn{1}{c|}{77.83} & 90.34 & \multicolumn{1}{c|}{47.22} & 84.01 & \multicolumn{1}{c|}{33.51} & 71.47 & \multicolumn{1}{c|}{26.20} & 56.72 \\ \midrule
OrganCMNIST & 89.02 & \multicolumn{1}{c|}{78.58} & 89.97 & \multicolumn{1}{c|}{56.71} & 89.64 & \multicolumn{1}{c|}{29.94} & 82.00 & \multicolumn{1}{c|}{23.11} & 69.80 & \multicolumn{1}{c|}{22.35} & 59.24 \\ \midrule
OrganSMNIST & 73.39 & \multicolumn{1}{c|}{68.18} & 68.00 & \multicolumn{1}{c|}{54.41} & 65.95 & \multicolumn{1}{c|}{22.23} & 63.32 & \multicolumn{1}{c|}{5.71} & 60.37 & \multicolumn{1}{c|}{5.10} & 54.12 \\ \bottomrule
\end{tabular}%
}
\end{table*}

\begin{table}[]
\centering
\caption{Model accuracy (\%) under Random Bit Flip faults across MedMNIST datasets, where 
k denotes the number of random bit flips applied.}
\label{rbitflip}
\resizebox{0.9\columnwidth}{!}{%
\begin{tabular}{|c|c|c|c|c|c|c|}
\hline
\rowcolor[HTML]{EFEFEF} 
Dataset & Clean Acc & k=1 & k=2 & k=4 & k=8 & k=16 \\ \midrule
PathMNIST & 89.72 & 81.73 & 76.19 & 72.25 & 52.04 & 32.32 \\ \midrule
DermaMNIST & 79.35 & 78.64 & 75.15 & 77.96 & 48.10 & 34.42 \\ \midrule
OCTMNIST & 91.80 & 85.08 & 79.63 & 76.04 & 45.22 & 43.33 \\ \midrule
PneumoniaMNIST & 93.27 & 90.11 & 87.04 & 87.07 & 66.80 & 59.94 \\ \midrule
RetinaMNIST & 51.00 & 49.00 & 50.70 & 49.80 & 41.03 & 42.21 \\ \midrule
BreastMNIST & 88.46 & 87.08 & 85.55 & 85.54 & 56.19 & 54.52 \\ \midrule
BloodMNIST & 97.63 & 93.27 & 84.54 & 79.55 & 53.46 & 28.43 \\ \hline
OrganAMNIST & 89.19 & 81.36 & 79.43 & 78.13 & 43.27 & 31.67 \\ \midrule
OrganCMNIST & 89.02 & 81.15 & 73.19 & 72.85 & 40.50 & 26.00 \\ \midrule
OrganSMNIST & 73.39 & 67.00 & 63.73 & 61.56 & 38.99 & 24.96 \\ \bottomrule
\end{tabular}%
}
\end{table}

\begin{table*}[]
\renewcommand{\arraystretch}{3}
\centering
\caption{Accuracy (\%) under layer-wise bit-flip faults for different values of k.}
\label{lbitflip}
\resizebox{1.05\textwidth}{!}{%
\begin{tabular}{|c|c|cllll|cllll|cllll|cllll|cllll|lllll|lllll|}
\hline
\rowcolor[HTML]{EFEFEF} 
\cellcolor[HTML]{EFEFEF} & \cellcolor[HTML]{EFEFEF} & \multicolumn{5}{c|}{\cellcolor[HTML]{EFEFEF}Patch\_embed} & \multicolumn{5}{c|}{\cellcolor[HTML]{EFEFEF}stage0\_layers.0} & \multicolumn{5}{c|}{\cellcolor[HTML]{EFEFEF}stage1\_layers.1} & \multicolumn{5}{c|}{\cellcolor[HTML]{EFEFEF}stage2\_layers.2} & \multicolumn{5}{c|}{\cellcolor[HTML]{EFEFEF}stage3\_layers.3} & \multicolumn{5}{c|}{\cellcolor[HTML]{EFEFEF}classifier\_head} & \multicolumn{5}{c|}{\cellcolor[HTML]{EFEFEF}ssm\_related} \\ \cline{3-37} 
\rowcolor[HTML]{EFEFEF} 
\multirow{-2}{*}{\cellcolor[HTML]{EFEFEF}Dataset} & \multirow{-2}{*}{\cellcolor[HTML]{EFEFEF}Clean Acc} & \multicolumn{1}{c|}{\cellcolor[HTML]{EFEFEF}1} & \multicolumn{1}{c|}{\cellcolor[HTML]{EFEFEF}2} & \multicolumn{1}{c|}{\cellcolor[HTML]{EFEFEF}4} & \multicolumn{1}{c|}{\cellcolor[HTML]{EFEFEF}8} & \multicolumn{1}{c|}{\cellcolor[HTML]{EFEFEF}16} & \multicolumn{1}{c|}{\cellcolor[HTML]{EFEFEF}1} & \multicolumn{1}{c|}{\cellcolor[HTML]{EFEFEF}2} & \multicolumn{1}{c|}{\cellcolor[HTML]{EFEFEF}4} & \multicolumn{1}{c|}{\cellcolor[HTML]{EFEFEF}8} & \multicolumn{1}{c|}{\cellcolor[HTML]{EFEFEF}16} & \multicolumn{1}{c|}{\cellcolor[HTML]{EFEFEF}1} & \multicolumn{1}{c|}{\cellcolor[HTML]{EFEFEF}2} & \multicolumn{1}{c|}{\cellcolor[HTML]{EFEFEF}4} & \multicolumn{1}{c|}{\cellcolor[HTML]{EFEFEF}8} & \multicolumn{1}{c|}{\cellcolor[HTML]{EFEFEF}16} & \multicolumn{1}{c|}{\cellcolor[HTML]{EFEFEF}1} & \multicolumn{1}{c|}{\cellcolor[HTML]{EFEFEF}2} & \multicolumn{1}{c|}{\cellcolor[HTML]{EFEFEF}4} & \multicolumn{1}{c|}{\cellcolor[HTML]{EFEFEF}8} & \multicolumn{1}{c|}{\cellcolor[HTML]{EFEFEF}16} & \multicolumn{1}{c|}{\cellcolor[HTML]{EFEFEF}1} & \multicolumn{1}{c|}{\cellcolor[HTML]{EFEFEF}2} & \multicolumn{1}{c|}{\cellcolor[HTML]{EFEFEF}4} & \multicolumn{1}{c|}{\cellcolor[HTML]{EFEFEF}8} & \multicolumn{1}{c|}{\cellcolor[HTML]{EFEFEF}16} & \multicolumn{1}{c|}{\cellcolor[HTML]{EFEFEF}1} & \multicolumn{1}{c|}{\cellcolor[HTML]{EFEFEF}2} & \multicolumn{1}{c|}{\cellcolor[HTML]{EFEFEF}4} & \multicolumn{1}{c|}{\cellcolor[HTML]{EFEFEF}8} & \multicolumn{1}{c|}{\cellcolor[HTML]{EFEFEF}16} & \multicolumn{1}{c|}{\cellcolor[HTML]{EFEFEF}1} & \multicolumn{1}{c|}{\cellcolor[HTML]{EFEFEF}2} & \multicolumn{1}{c|}{\cellcolor[HTML]{EFEFEF}4} & \multicolumn{1}{c|}{\cellcolor[HTML]{EFEFEF}8} & \multicolumn{1}{c|}{\cellcolor[HTML]{EFEFEF}16} \\ \midrule

PathMNIST & 89.72 & \multicolumn{1}{c|}{80.23} & \multicolumn{1}{c|}{67.44} & \multicolumn{1}{c|}{52.19} & \multicolumn{1}{c|}{22.02} & \multicolumn{1}{c|}{27.48} & \multicolumn{1}{c|}{82.58} & \multicolumn{1}{c|}{68.30} & \multicolumn{1}{c|}{64.82} & \multicolumn{1}{c|}{32.81} & \multicolumn{1}{c|}{22.19} & \multicolumn{1}{c|}{82.60} & \multicolumn{1}{c|}{68.39} & \multicolumn{1}{c|}{64.80} & \multicolumn{1}{c|}{43.13} & \multicolumn{1}{c|}{25.41} & \multicolumn{1}{c|}{82.62} & \multicolumn{1}{c|}{68.40} & \multicolumn{1}{c|}{64.86} & \multicolumn{1}{c|}{21.33} & \multicolumn{1}{c|}{25.75} & \multicolumn{1}{c|}{86.18} & \multicolumn{1}{c|}{77.53} & \multicolumn{1}{c|}{69.89} & \multicolumn{1}{c|}{44.27} & \multicolumn{1}{c|}{24.12} & \multicolumn{1}{c|}{86.26} & \multicolumn{1}{c|}{75.23} & \multicolumn{1}{c|}{70.13} & \multicolumn{1}{c|}{45.18} & \multicolumn{1}{c|}{36.74} & \multicolumn{1}{c|}{81.76} & \multicolumn{1}{c|}{66.34} & \multicolumn{1}{c|}{56.22} & \multicolumn{1}{c|}{28.06} & \multicolumn{1}{c|}{19.57} \\ \midrule

DermaMNIST & 79.35 & \multicolumn{1}{c|}{71.40} & \multicolumn{1}{l|}{56.01} & \multicolumn{1}{l|}{44.62} & \multicolumn{1}{l|}{7.10} & 13.52 & \multicolumn{1}{c|}{71.73} & \multicolumn{1}{l|}{56.50} & \multicolumn{1}{l|}{52.70} & \multicolumn{1}{l|}{18.51} & 7.07 & \multicolumn{1}{c|}{71.76} & \multicolumn{1}{l|}{56.33} & \multicolumn{1}{l|}{52.55} & \multicolumn{1}{l|}{26.13} & 7.52 & \multicolumn{1}{c|}{71.79} & \multicolumn{1}{l|}{56.55} & \multicolumn{1}{l|}{52.75} & \multicolumn{1}{l|}{7.27} & 10.91 & \multicolumn{1}{c|}{72.51} & \multicolumn{1}{l|}{68.31} & \multicolumn{1}{l|}{59.59} & \multicolumn{1}{l|}{32.64} & 21.92 & \multicolumn{1}{l|}{73.37} & \multicolumn{1}{l|}{67.33} & \multicolumn{1}{l|}{64.74} & \multicolumn{1}{l|}{43.27} & 42.76 & \multicolumn{1}{l|}{72.37} & \multicolumn{1}{l|}{56.97} & \multicolumn{1}{l|}{46.12} & \multicolumn{1}{l|}{24.11} & 8.13 \\ \midrule

OCTMNIST & 91.80 & \multicolumn{1}{c|}{83.05} & \multicolumn{1}{l|}{71.18} & \multicolumn{1}{l|}{58.46} & \multicolumn{1}{l|}{28.37} & 35.05 & \multicolumn{1}{c|}{85.19} & \multicolumn{1}{l|}{71.79} & \multicolumn{1}{l|}{68.40} & \multicolumn{1}{l|}{38.36} & 28.34 & \multicolumn{1}{c|}{85.12} & \multicolumn{1}{l|}{71.79} & \multicolumn{1}{l|}{68.46} & \multicolumn{1}{l|}{45.03} & 31.68 & \multicolumn{1}{c|}{85.13} & \multicolumn{1}{l|}{71.77} & \multicolumn{1}{l|}{68.44} & \multicolumn{1}{l|}{28.67} & 31.68 & \multicolumn{1}{c|}{91.81} & \multicolumn{1}{l|}{81.79} & \multicolumn{1}{l|}{65.41} & \multicolumn{1}{l|}{49.58} & 30.58 & \multicolumn{1}{l|}{89.36} & \multicolumn{1}{l|}{78.07} & \multicolumn{1}{l|}{70.04} & \multicolumn{1}{l|}{55.03} & 47.24 & \multicolumn{1}{l|}{85.77} & \multicolumn{1}{l|}{73.66} & \multicolumn{1}{l|}{61.60} & \multicolumn{1}{l|}{40.17} & 27.58 \\ \midrule

PneumoniaMNIST & 93.27 & \multicolumn{1}{c|}{87.44} & \multicolumn{1}{l|}{72.60} & \multicolumn{1}{l|}{68.45} & \multicolumn{1}{l|}{40.32} & 45.68 & \multicolumn{1}{c|}{87.72} & \multicolumn{1}{l|}{76.52} & \multicolumn{1}{l|}{73.75} & \multicolumn{1}{l|}{48.61} & 40.29 & \multicolumn{1}{c|}{87.80} & \multicolumn{1}{l|}{76.59} & \multicolumn{1}{l|}{73.79} & \multicolumn{1}{l|}{54.24} & 40.30 & \multicolumn{1}{c|}{87.70} & \multicolumn{1}{l|}{76.54} & \multicolumn{1}{l|}{73.74} & \multicolumn{1}{l|}{40.72} & 43.10 & \multicolumn{1}{c|}{88.94} & \multicolumn{1}{l|}{88.94} & \multicolumn{1}{l|}{78.74} & \multicolumn{1}{l|}{67.09} & 48.32 & \multicolumn{1}{l|}{89.61} & \multicolumn{1}{l|}{85.55} & \multicolumn{1}{l|}{76.39} & \multicolumn{1}{l|}{52.13} & 62.85 & \multicolumn{1}{l|}{89.65} & \multicolumn{1}{l|}{76.77} & \multicolumn{1}{l|}{75.99} & \multicolumn{1}{l|}{54.78} & 44.06 \\ \midrule

RetinaMNIST & 51.00 & \multicolumn{1}{c|}{49.85} & \multicolumn{1}{l|}{48.13} & \multicolumn{1}{l|}{47.00} & \multicolumn{1}{l|}{43.80} & 44.44 & \multicolumn{1}{c|}{50.25} & \multicolumn{1}{l|}{48.74} & \multicolumn{1}{l|}{48.39} & \multicolumn{1}{l|}{45.03} & 43.90 & \multicolumn{1}{c|}{50.25} & \multicolumn{1}{l|}{48.75} & \multicolumn{1}{l|}{46.36} & \multicolumn{1}{l|}{45.75} & 44.28 & \multicolumn{1}{c|}{50.25} & \multicolumn{1}{l|}{48.75} & \multicolumn{1}{l|}{48.38} & \multicolumn{1}{l|}{43.88} & 44.26 & \multicolumn{1}{c|}{51.00} & \multicolumn{1}{l|}{46.36} & \multicolumn{1}{l|}{45.05} & \multicolumn{1}{l|}{38.51} & 37.61 & \multicolumn{1}{l|}{50.66} & \multicolumn{1}{l|}{50.33} & \multicolumn{1}{l|}{44.80} & \multicolumn{1}{l|}{31.45} & 27.05 & \multicolumn{1}{l|}{47.15} & \multicolumn{1}{l|}{40.99} & \multicolumn{1}{l|}{43.03} & \multicolumn{1}{l|}{28.29} & 36.81 \\ \midrule

BreastMNIST & 88.46 & \multicolumn{1}{c|}{80.64} & \multicolumn{1}{l|}{69.87} & \multicolumn{1}{l|}{57.79} & \multicolumn{1}{l|}{29.94} & 34.52 & \multicolumn{1}{c|}{82.34} & \multicolumn{1}{l|}{69.97} & \multicolumn{1}{l|}{66.99} & \multicolumn{1}{l|}{39.20} & 30.00 & \multicolumn{1}{c|}{82.31} & \multicolumn{1}{l|}{70.00} & \multicolumn{1}{l|}{66.99} & \multicolumn{1}{l|}{48.56} & 32.34 & \multicolumn{1}{c|}{82.31} & \multicolumn{1}{l|}{70.00} & \multicolumn{1}{l|}{66.92} & \multicolumn{1}{l|}{30.32} & 33.11 & \multicolumn{1}{c|}{88.46} & \multicolumn{1}{l|}{75.38} & \multicolumn{1}{l|}{76.15} & \multicolumn{1}{l|}{57.56} & 32.60 & \multicolumn{1}{c|}{82.21} & \multicolumn{1}{l|}{82.76} & \multicolumn{1}{l|}{72.08} & \multicolumn{1}{l|}{51.06} & 56.22 & \multicolumn{1}{l|}{85.42} & \multicolumn{1}{l|}{82.21} & \multicolumn{1}{l|}{65.86} & \multicolumn{1}{l|}{50.22} & 38.08 \\ \midrule

BloodMNIST & 97.63 & \multicolumn{1}{c|}{86.87} & \multicolumn{1}{l|}{65.08} & \multicolumn{1}{l|}{51.43} & \multicolumn{1}{l|}{11.66} &  \multicolumn{1}{l|}{15.27} & \multicolumn{1}{c|}{88.55} & \multicolumn{1}{l|}{70.43} & \multicolumn{1}{l|}{65.92} & \multicolumn{1}{l|}{25.23} & \multicolumn{1}{c|}{11.66} & \multicolumn{1}{c|}{88.58} & \multicolumn{1}{l|}{70.48} & \multicolumn{1}{l|}{65.95} & \multicolumn{1}{l|}{38.90} & \multicolumn{1}{l|}{11.76} & \multicolumn{1}{c|}{88.58} & \multicolumn{1}{l|}{70.48} & \multicolumn{1}{l|}{65.96} & \multicolumn{1}{l|}{11.99} & \multicolumn{1}{l|}{16.18} & \multicolumn{1}{c|}{88.64} & \multicolumn{1}{l|}{89.17} & \multicolumn{1}{l|}{62.88} & \multicolumn{1}{l|}{44.97} & \multicolumn{1}{l|}{19.03} & \multicolumn{1}{l|}{91.43} & \multicolumn{1}{l|}{88.36} & \multicolumn{1}{l|}{78.79} & \multicolumn{1}{l|}{46.73} & \multicolumn{1}{l|}{44.41} & \multicolumn{1}{l|}{89.40} & \multicolumn{1}{l|}{72.67} & \multicolumn{1}{l|}{57.92} & \multicolumn{1}{l|}{30.50} & \multicolumn{1}{l|}{14.33} \\ \midrule

OrganAMNIST & 89.19 & \multicolumn{1}{c|}{80.82} & \multicolumn{1}{l|}{62.96} & \multicolumn{1}{l|}{51.41} & \multicolumn{1}{l|}{9.99} & 18.17 & \multicolumn{1}{c|}{80.86} & \multicolumn{1}{l|}{64.10} & \multicolumn{1}{l|}{59.93} & \multicolumn{1}{l|}{22.49} & 9.98 & \multicolumn{1}{c|}{80.88} & \multicolumn{1}{l|}{64.20} & \multicolumn{1}{l|}{60.02} & \multicolumn{1}{l|}{35.01} & 14.08 & \multicolumn{1}{c|}{80.86} & \multicolumn{1}{l|}{64.18} & \multicolumn{1}{l|}{60.02} & \multicolumn{1}{l|}{10.13} & 14.17 & \multicolumn{1}{c|}{85.25} & \multicolumn{1}{l|}{72.57} & \multicolumn{1}{l|}{68.54} & \multicolumn{1}{l|}{31.70} & 15.55 & \multicolumn{1}{l|}{86.08} & \multicolumn{1}{l|}{79.65} & \multicolumn{1}{l|}{67.10} & \multicolumn{1}{l|}{51.48} & 36.79 & \multicolumn{1}{l|}{80.86} & \multicolumn{1}{l|}{64.81} & \multicolumn{1}{l|}{52.02} & \multicolumn{1}{l|}{23.32} & 10.52 \\ \midrule

OrganCMNIST & 89.02 & \multicolumn{1}{c|}{80.44} & \multicolumn{1}{l|}{64.76} & \multicolumn{1}{l|}{52.81} & \multicolumn{1}{l|}{14.04} & 21.01 & \multicolumn{1}{c|}{81.08} & \multicolumn{1}{l|}{65.28} & \multicolumn{1}{l|}{61.40} & \multicolumn{1}{l|}{25.87} & 14.03 & \multicolumn{1}{c|}{81.12} & \multicolumn{1}{l|}{65.35} & \multicolumn{1}{l|}{61.40} & \multicolumn{1}{l|}{33.93} & 18.00 & \multicolumn{1}{c|}{81.12} & \multicolumn{1}{l|}{65.34} & \multicolumn{1}{l|}{61.40} & \multicolumn{1}{l|}{13.63} & 17.97 & \multicolumn{1}{c|}{80.88} & \multicolumn{1}{l|}{72.93} & \multicolumn{1}{l|}{65.45} & \multicolumn{1}{l|}{38.20} & 15.26 & \multicolumn{1}{l|}{85.75} & \multicolumn{1}{l|}{77.31} & \multicolumn{1}{l|}{70.06} & \multicolumn{1}{l|}{51.46} & 32.19 & \multicolumn{1}{l|}{80.95} & \multicolumn{1}{l|}{64.92} & \multicolumn{1}{l|}{52.77} & \multicolumn{1}{l|}{26.73} & 13.76 \\ \midrule

OrganSMNIST & 73.39 & \multicolumn{1}{c|}{65.23} & \multicolumn{1}{l|}{50.59} & \multicolumn{1}{l|}{44.02} & \multicolumn{1}{l|}{12.43} & 17.33 & \multicolumn{1}{c|}{66.81} & \multicolumn{1}{l|}{54.08} & \multicolumn{1}{l|}{50.83} & \multicolumn{1}{l|}{22.07} & 12.36 & \multicolumn{1}{c|}{66.97} & \multicolumn{1}{l|}{54.12} & \multicolumn{1}{l|}{50.89} & \multicolumn{1}{l|}{31.63} & 12.48 & \multicolumn{1}{c|}{66.96} & \multicolumn{1}{l|}{54.11} & \multicolumn{1}{l|}{50.91} & \multicolumn{1}{l|}{12.39} & 15.61 & \multicolumn{1}{c|}{70.10} & \multicolumn{1}{l|}{56.89} & \multicolumn{1}{l|}{51.85} & \multicolumn{1}{l|}{35.94} & 14.60 & \multicolumn{1}{l|}{71.52} & \multicolumn{1}{l|}{67.11} & \multicolumn{1}{l|}{62.89} & \multicolumn{1}{l|}{40.33} & 29.03 & \multicolumn{1}{l|}{66.83} & \multicolumn{1}{l|}{53.45} & \multicolumn{1}{l|}{44.24} & \multicolumn{1}{l|}{21.43} & 12.32 \\ \bottomrule
\end{tabular}%
}
\end{table*}

\section{Experiments}

This section describes the experimental setup and summarizes the main results. We adopt \textbf{MedMamba}~\cite{yue2024medmamba} as our base architecture and evaluate on \textbf{MedMNIST}~\cite{medmnist_website}, a collection of standardized, low-resolution medical imaging benchmarks that enables consistent comparisons across tasks and modalities.

\subsection{Datasets}
We conduct experiments on MedMNIST~\cite{medmnist_website} using the official train/validation/test splits. Depending on the subset, the task is either multi-class classification or multi-label classification. Unless otherwise stated, we follow the dataset’s default preprocessing and label definitions to ensure reproducibility and fair comparison.

\subsection{Model and Training Protocol}
We use MedMamba~\cite{yue2024medmamba} as the backbone model. For each dataset, we train the model on the training split and select checkpoints based on validation performance. We report final performance on the held-out test split. All experiments use the same training recipe across datasets as much as possible (optimizer, learning-rate schedule, and regularization), with only minimal task-specific adjustments (e.g., number of classes). We report the accuracy of the model on the test set as the performance metric.

\subsection{Robustness Evaluation: Med-Mamba-Adv}

Table \ref{Wattack} summarizes the white-box attack results under FGSM and PGD ($\epsilon = 1/255$, 20 steps). The model shows strong clean performance on most datasets (e.g., BloodMNIST: 97.63\%), but adversarial perturbations substantially reduce accuracy across all tasks. In particular, PGD causes the most severe degradation, with some datasets collapsing to near-random performance or even 0\% accuracy (e.g., PneumoniaMNIST).

This behavior is expected since PGD is a stronger iterative attack that repeatedly refines the perturbation, whereas FGSM applies a single-step update and is therefore less effective. The large gap between clean and adversarial accuracy indicates that the model is highly sensitive to small input perturbations and lacks robustness under strong white-box attacks. Additionally, robustness varies across datasets: BloodMNIST and the OrganMNIST subsets retain relatively higher adversarial accuracy compared to others, suggesting that dataset characteristics such as class separability, texture patterns, and image complexity may influence adversarial vulnerability.

Table \ref{drop} reports the PatchDrop evaluation results, showing how the model’s classification accuracy changes as an increasing percentage of image patches are removed. When no patches are dropped (0\%), the model achieves its baseline clean performance, while higher drop ratios progressively degrade accuracy across most MedMNIST datasets. The results demonstrate that the model is sensitive to missing regions, particularly at high dropout levels (e.g., 50\% and 56.25\%). Some datasets show a gradual decline in performance such as PathMNIST and the OrganMNIST subsets, suggesting that the model can still exploit remaining global structure when partial information is available. In contrast, datasets like BloodMNIST experience a sharp drop in accuracy even at moderate patch removal implying that the classification relies heavily on fine-grained local details that are easily destroyed by patch removal.

Table \ref{noise} presents the robustness evaluation under common image corruptions, namely Gaussian noise (N) and defocus blur (B), across five severity (intensity) levels. The results report classification accuracy for each MedMNIST dataset under clean conditions and under progressively stronger corruptions. Overall, both noise and blur degrade performance as the intensity increases, confirming that the model is sensitive to distribution shifts that distort pixel-level information or reduce image sharpness.

\subsection{Robustness Evaluation: Med-Mamba-Hammer}
Table \ref{rbitflip} summarizes the impact of Random Bit Flip perturbations on the model’s performance across MedMNIST datasets. We report the clean accuracy alongside the accuracy obtained after injecting bit-flip faults at different levels (k=1,2,4,8,16), where larger values of k indicate more severe corruption. Overall, these results indicate that increasing hardware-level faults can significantly reduce inference reliability, highlighting the need for fault-aware or fault-tolerant deployment.

Table \ref{lbitflip} presents a layer-wise bit-flip sensitivity analysis, where random bit flips are injected into one model component at a time (Patch\_embed, stage layers, classifier head, and SSM-related modules). For each dataset, we report the clean accuracy and the accuracy after injecting k={1,2,4,8,16} bit flips in the selected layer. Overall, the results show that fault impact depends strongly on which layer is corrupted, and accuracy generally decreases as the number of bit flips increases. A consistent trend is that early feature extraction modules (especially Patch\_embed) and SSM-related components often lead to a larger performance drop, since corruption at these stages propagates through the network and affects all downstream representations.

The worst-case bit-flip experiment reveals that MedMamba is extremely vulnerable to rare but highly damaging exponent-bit corruptions: although the clean model achieves 97.63\% test accuracy on BloodMNIST, a worst-case search over corrupted candidates shows that flipping only one exponent bit (K=1) can collapse performance to 9.09\% accuracy. Increasing the corruption budget further degrades performance, reaching 4.65\% accuracy at K=8 and 0.67\% at K=16.

\section{Conclusion and Future Work}

This paper evaluated the reliability of Mamba-based medical imaging models under both input-level perturbations and hardware-style bit-flip faults. Across MedMNIST datasets, MedMamba achieves strong clean accuracy but shows substantial vulnerability to white-box attacks (FGSM/PGD), PatchDrop occlusions, and common corruptions such as Gaussian noise and defocus blur. In the hardware threat model, random and layer-wise bit flips degrade performance with increasing fault budgets, and the worst-case setting reveals that even a single high-impact exponent-bit flip can collapse accuracy to near-random levels.
Future work includes improving robustness via adversarial/corruption-aware training, applying fault-tolerant techniques such as selective protection of sensitive modules and lightweight error detection/correction, and extending evaluation to higher-resolution clinical datasets and additional medical imaging tasks.

\bibliographystyle{ieeetr}
\bibliography{ref.bib}

\end{document}